\documentclass[12pt]{elsarticle}
\usepackage{relsize}
\usepackage{multirow}
\usepackage{array,multirow}
\usepackage{natbib}
\usepackage{amssymb}
\usepackage{amsmath}
\usepackage{graphicx}
\usepackage{bm}
\usepackage{epsfig}
\usepackage{textcomp}
\usepackage{wrapfig}
\usepackage{float}

\begin{document}
\begin{frontmatter}
\title{Chi-square goodness of fit tests for weighted histograms. Review and improvements.}
\author{N.D. Gagunashvili\corref{cor1}\fnref{fn1}}
\fntext[fn1]{Present address: Max-Planck-Institut f\"{u}r Kernphysik, PO Box
103980, \\ 69029 Heidelberg, Germany}
\ead{nikolai@unak.is}
\cortext[cor1]{Tel.: +354-4608505; fax: +354-4608998}
\address{University of Akureyri, Borgir, v/Nordursl\'od, IS-600 Akureyri, Iceland}
\begin{abstract}
Weighted histograms are used for the estimation of  probability density functions.
Computer simulation is the main domain of application of this type of histogram. A review of chi-square goodness of fit tests for weighted histograms is presented in this paper. Improvements are proposed to  these tests that have  size more close to its nominal value. Numerical examples are presented in this paper for evaluation of tests and  to demonstrate various applications of tests.
\end{abstract}
\end{frontmatter}
\section{Introduction}
A histogram with $m$ bins for a given probability density function (PDF)  $p(x)$ is used to estimate the probabilities
\begin{equation}
p_i=\int_{S_i}p(x)dx, \; i=1,\ldots ,m \label{p1}
\end{equation}
 that a random event  belongs to  bin $i$. Integration in (\ref{p1}) is done over the bin  $S_i$.

A histogram can be obtained as a result of a random experiment with PDF $p(x)$.
Let us denote the number of random events belonging to the $i$th bin
of the  histogram as $n_{i}$. The total number of events $n$ in the histogram is equal to
 \begin{equation}
 n=\sum_{i=1}^{m}{n_i}.
 \end{equation}
   The quantity
   \begin{equation}
    \hat{p}_i= n_{i}/n
    \end{equation}
     is an estimator
of probability $p_i$  with expectation value
\begin{equation}
 \textrm E\,[\hat{p_i}]=p_i.
 \end{equation}
  The distribution
of the number of events for  bins of the  histogram is the multinomial distribution \cite{kendall}  and the probability of the random vector
$(n_1,\ldots ,n_m)$  is
\begin{equation}
P(n_1,\ldots ,n_m)=\frac{n!}{n_1!n_2! \ldots n_m!} \; p_1^{n_1}
\ldots p_m^{n_m},\text{ } \sum_{i=1}^{m} p_i=1.
\end{equation}

 A weighted histogram or a histogram of weighted events is used again for estimating the probabilities $p_i$ (\ref{p1}), see Ref. \cite{gagunash}. It  is obtained as a result of a random experiment with probability density function $g(x)$ that generally does not coincide with PDF $p(x)$.
The sum of weights of events for  bin $i$ is defined as:
 \begin{equation}
W_i=  \sum_{k=1}^{n_i}w_i(k), \label{ffffweight}
\end{equation}
where $n_i$ is the number of events at bin $i$ and $w_i(k)$ is the weight of the $k$th event in the $i$th bin.  The statistic
\begin{equation}
\hat{p_i}=W_i/n  \label{west}
\end{equation}
is used to estimate $p_i$, where $n=\sum_{i=1}^{m}{n_i}$ is the total number of events for the histogram with  $m$ bins. Weights of events are chosen in such a way that the estimate  (\ref {west}) is unbiased,
\begin{equation}
 \textrm E [\hat{p_i}]=p_i.
 \end{equation}
The usual histogram  is a weighted histogram with weights of events equal to 1.

The two  examples of weighted histograms are considered below:\\

\subsection{Example 1}
To define a weighted histogram let us write the probability $p_i$
(\ref{p1}) for a given PDF  $p(x)$  in the form
\begin{equation}
p_i= \int_{S_i}p(x)dx = \int_{S_i}w(x)g(x)dx, \label{weightg}
\end{equation}
where
\begin{equation}
w(x)=p(x)/g(x) \label{fweightg}
\end{equation}
 is the weight function and $g(x)$ is some other probability density function. The function $g(x)$ must be $>0$ for points $x$, where $p(x)\neq 0$. The weight $w(x)=0$ if $p(x)=0$, see Ref.  \cite{sobol}.

The weighted histogram is obtained from a random experiment with a probability density function $g(x)$, and the weights of the events are calculated according to (\ref{fweightg}).\\

\subsection{Example 2}
The probability density function  $p_{rec}(x)$ of a reconstructed
characteristic $x$ of an event obtained from a detector with
finite resolution and limited acceptance can be represented as
\begin{equation}
p_{rec}(x) \propto \int_{\Omega'} p_{tr}(x')A(x')R(x|x') \,dx',
\label{p1_main}
\end{equation}
where $p_{tr}(x')$ is the true PDF, $A(x')$ is the acceptance of the
setup, i.e. the probability of recording an event with a
characteristic $x'$, and $R(x|x')$ is the experimental resolution,
i.e. the probability of obtaining $x$ instead of $x'$ after the
reconstruction of the event. The integration in (\ref{p1_main}) is
carried out over the domain $\Omega'$ of the variable $x'$.
Total probability that  an event will not be registered is equal to
\begin{equation}
\overline{p}=  \int_{\Omega'} p_{tr}(x')(1-A(x')) \,dx'.
\label{p1_main8}
\end{equation}
The  sum of probabilities
\begin{equation}
 \int_{\Omega} \int_{\Omega'}p_{tr}(x')A(x')R(x|x') \,dx'dx+\int_{\Omega'} p_{tr}(x')(1-A(x')) \,dx'=1
\label{p1_main3}
\end{equation}
because
\begin{equation}
 \int_{\Omega}\int_{\Omega'} p_{tr}(x')A(x')R(x|x') \,dx'dx=\int_{\Omega'} p_{tr}(x')A(x'), \,dx',
\end{equation}
where $\Omega$ domain of the variable $x$.

A histogram of the  PDF $p_{rec}(x)$ can be obtained as a result of a random experiment (simulation) that has three steps \cite{sobol}:
\begin{enumerate}
\item A random value $x'$ is chosen according to a PDF $p_{tr}(x')$.
\item We go back to step 1 again
with probability $1-A(x')$, and to step 3 with probability $A(x')$.
\item A random value $x$ is chosen according to the PDF $R(x|x')$.
\end{enumerate}

The quantity $\hat {p_i}= n_{i}/n$, where $n_{i}$ is
the number of events belonging to the $i$th bin for a histogram
with total number of events $n$ in random experiment (at step 1), is an estimator of
$p_{i}$,
\begin{equation}
p_{i}= \int_{S_i} \int_{\Omega'}p_{tr}(x')A(x')R(x|x') \, dx' \, dx,\; i=1,\ldots ,m,
\label{p2i}
\end{equation}
with the expectation value of the estimator
\begin{equation}
\textrm E \,[\hat{p_{i}}]=p_{i}.
\end{equation}
The quantity $\hat{\overline{p}}= \overline{n}/n$, where $\overline{n}$ is
the number of events that were lost, is an estimator of $\overline{p}$ (\ref{p1_main8})
with the expectation value of the estimator
\begin{equation}
\textrm E \,[\hat{\overline{p}}]= \overline{p}.
\end{equation}
Notice that
\begin{equation}
\sum_{i=1}^{m} p_i+\overline{p}=1 \,\, \text{and} \,\, \sum_{i=1}^{m} n_i+\overline{n}=n.
\end{equation}
In experimental particle and nuclear physics, step 3 is the most
time-consuming step of the Monte Carlo simulation. This step is
related to the simulation of the process of transport of particles
through a medium and the rather complex registration apparatus.

To use the results of the simulation with some  PDF $g_{tr}(x')$ for calculating
a weighted histogram of events with a true PDF $p_{tr}(x')$, we write the equation for
$p_{i}$ in the form

\begin{equation}
p_{i}=  \int_{S_i} \int_{\Omega'} w(x')g_{tr}(x')A(x')R(x|x') \, dx' \, dx,
\label{p23}
\end{equation}
where
\begin{equation}
w(x')=p_{tr}(x')/g_{tr}(x') \label{fweight}
\end{equation}
is the weight function.

The weighted histogram for the PDF $p_{rec}(x)$ can be obtained using events with reconstructed characteristic $x$ and weights calculated according to
(\ref{fweight}).

In this way, we avoid step 3 of the simulation
procedure, which is important in cases where one needs to
calculate Monte Carlo reconstructed histograms for many different
true PDFs.

The probability that  an event will not be registered can be represented as
\begin{equation}
\overline{p}=  \int_{\Omega'} w(x')g_{tr}(x')(1-A(x')) \,dx',
\label{p1_main2}
\end{equation}
and is  estimated the same way using events with weights calculated according formula  (\ref{fweight}).


\section{Goodness of fit tests}
The problem of goodness of fit is to test the hypothesis
\begin{equation}
H_0: p_1=p_{10},\ldots, p_{m-1}=p_{m-1,0} \text{  vs.  } H_a: p_i \neq p_{i0} \text{  for some  } i,
\end{equation}
where $p_{i0}$ are specified probabilities, and $\sum_{i=1}^{m} p_{i0}=1$. The test is used in a data  analysis for comparing theoretical frequencies $np_{i0}$ with observed frequencies $n_i$. This classical problem remains of current practical interest. The test statistic for a histogram with unweighted entries
\begin{equation}
X^2=\sum_{i=1}^{m} \frac{(n_i-np_{i0})^2}{np_{i0}} \label{basic11}
\end{equation}
was suggested by Pearson \cite{pearson}. Pearson showed that the statistic (\ref{basic11}) has  approximately a $\chi^2_{m-1}$ distribution if the hypothesis $H_0$ is true.

\subsection{The contemporary  proof of Pearson's result}

The expectation values of the observed frequency $n_i$, if hypothesis $H_0$ is valid, equal to:
\begin{equation}
\textrm E [n_i]=np_{i0}, \; i=1,\ldots,m \label{ew}
\end{equation}
and its covariance matrix $\mathbf{\Gamma}$ has elements:
\hspace*{-4cm} $$\! \! \! \! \! \! \! \!\! \! \! \! \! \! \! \!\! \! \! \! \! \! \!\gamma_{ij}= \begin{cases}
     np_{i0}(1-p_{i0}) \text { for } i=j\\
     -np_{i0}p_{j0}   \text{\hspace *{0.7cm} for  } i \neq j
      \end{cases}$$
Notice that the covariance matrix $\mathbf{\Gamma}$ is singular \cite{kendall2}.

Let us now introduce the multivariate statistic
\begin{equation}
(\textbf{n}-n\textbf{p}_0)^t \mathbf{\Gamma_k^{-1}}(\textbf{n}-n\textbf{p}_0),\label{hot}
\end{equation}
where \\ $\textbf{n}=(n_1,\ldots, n_{k-1},n_{k+1},\ldots,n_{m})^t$,
$\textbf{p}_0=(p_{10},\ldots,p_{k-1,0},p_{k+1,0},\ldots,p_{m0})^t$
and $\mathbf{\Gamma_k}=(\gamma_{ij})_{(m-1)\times(m-1)}$ is the covariance
matrix for a histogram without bin $k$. The matrix $\mathbf{\Gamma_k}$ has the form
\begin{equation}
\mathbf{\Gamma_k}=n\,\textrm{diag}\,(p_{10}, \ldots
,p_{k-1,0},p_{k+1,0}, \ldots ,
p_{m 0})-n\textbf{p}_0\textbf{p}_0^t.
\end{equation}
The special form of this matrix permits one to find analytically
$\mathbf{\Gamma_k^{-1}}$  \cite{woodbury}:
\begin{equation}
\mathbf{\Gamma_k^{-1}}=\frac{1}{n}\textrm{diag}\,(\frac{1}{p_{10}}, \ldots
,\frac{1}{p_{k-1,0}},\frac{1}{p_{k+1,0}}, \ldots ,\frac{1}{p_{m 0}})+\frac{1}{np_{k,0}}\mathbf{\Theta},
\end{equation}
 where $\mathbf{\Theta}$ is $(m-1) \times (m-1)$ matrix with all elements unity.
 Finally the result of the calculation of expression (\ref{hot}) gives us the $X^2$ test statistic (\ref{basic11}).
Notice that the result  will be the same for any choice of  bin number $k$.

Asymptotically the vector $\textbf{n}$ has a normal distribution $\mathcal{N}(n
\textbf{p}_0,\mathbf{\Gamma_k^{1/2}})$, see Ref. \cite{kendall2}, and therefore the test statistic (\ref{basic11}) has $\chi^2_{m-1}$ distribution if hypothesis ${H_0}$ is true
\begin{equation}
 X^2 \sim \chi^2_{m-1}.
\end{equation}

\subsection{Generalization of the Pearson's chi-square test for weighted histograms}

The total sum of weights of events in $i$th  bin  $W_{i}$,
$i=1,\ldots,m$,  as proposed in  Ref. \cite{gagunash}, can be  considered as a sum of random variables
 \begin{equation}
W_i= \sum_{k=1}^{n_i}w_i(k), \label{ffffweight}
\end{equation}
where also the number of events  $n_i$ is a random value and the
weights $w_i(k),k=1,...,n_i$ are independent
 random variables with the same probability distribution function.
 The distribution of the number of events for bins of the histogram  is
  the multinomial distribution  and the probability of the random vector
  $(n_1,\ldots ,n_m)$ is
\begin{equation}
P(n_1,\ldots ,n_m)=\frac{n!}{n_1!n_2! \ldots n_m!} \; g_1^{n_1}
\ldots g_m^{n_m},\text{ } \sum_{i=1}^{m} g_i=1,
\end{equation}
where
$g_i$  is the probability that a random event  belongs to the  bin $i$.

 Let us denote the expectation values of the weights of events from the $i$th bin as
 \begin{equation}
 \textrm E[w_{i}]= \mu_i
 \end{equation}
  and the
  variances as
  \begin{equation}
  \textrm {Var}[w_{i}]= \sigma_i^2.
  \end{equation}
   The expectation value
  of the total sum of weights $W_{i}, i=1,\ldots,m$, see Ref. \cite{gnedenko},  is:
\begin{equation}
\textrm E[W_i]= \textrm E[\sum_{k=1}^{n_i}w_{i}(k)]=  \textrm E[
w_{i}] \textrm E[n_i] =  n\mu_ig_i.\label{eew}
\end{equation}
The diagonal elements $\gamma_{ii}$ of the  covariance matrix of the
vector $(W_1,\ldots ,W_m)$, see Ref. \cite{gnedenko}, are equal to
\begin{equation}
\gamma_{ii}=\sigma_i^2 g_in+\mu_i^2 g_i(1-g_i)n=
n\alpha_{2i}g_i-n\mu_i^2g_i^2, \label{giii}
\end{equation}
where
\begin{equation}
 \alpha_{2i}= \textrm E[w_{i}^2].
\end{equation}
  The non-diagonal elements
$\gamma_{ij},\,i\neq j$  are equal to:
\begin{equation}
\begin{split}
\gamma_{ij}=\sum_{k=0}^n\sum_{l=0}^{n}\textrm E
\,[\sum_{u=1}^k\sum_{v=1}^l w_i(u)w_j(v)]h(k,l)
-\textrm E[W_i]\textrm E[W_j]\\
=\sum_{k=0}^n\sum_{l=0}^{n}\textrm E [w_{i}w_{j}]h(k,l)kl-\mu_ing_i\mu_jng_j\quad\quad\quad\quad\,\\
=\mu_i\mu_j(-g_ig_jn+g_ig_jn^2 )-
\mu_ing_i\mu_jng_j\quad\quad\quad\quad\;\:\,\,\\=-n\mu_i\mu_j
g_ig_j,\quad\quad\quad\quad\quad\quad\quad\quad\quad\quad\quad\quad\quad\quad\quad\,
 \label{gij}
\end{split}
\end{equation}
where $h(k,l)$ is the probability that $k$ events belong to bin $i$
and $l$ events to bin $j$.

For weighted histograms again the problem of goodness of fit is to test the hypothesis
\begin{equation}
H_0: p_1=p_{10},\ldots ,p_{m-1}=p_{m-1,0} \text{  vs.  } H_a: p_i \neq p_{i0} \text{  for some  } i,
\end{equation}
where $p_{i0}$ are specified probabilities, and $\sum_{i=1}^{m} p_{i0}=1$.
 If hypothesis $H_0$ is true then
\begin{equation}
\textrm E[W_i]=n\mu_ig_i=np_{i0}, \; i=1,\ldots,m \label{ew}
\end{equation}
and
\begin{equation}
g_i=p_{i0}/\mu_i, \; i=1,\ldots,m. \label{pi}
\end{equation}
We can substitute $g_i$ to Eqs. (\ref{giii}) and (\ref{gij})  which gives
the covariance matrix $\mathbf{\Gamma}$ with elements:
\hspace*{-4cm} $$\! \! \! \! \! \! \! \!\! \! \! \! \! \! \! \!\! \! \! \! \! \! \!\gamma_{ij}= \begin{cases}
     np_{i0}( r_i^{-1} -p_{i0})\,\,\,
      \text { for } i=j\\
     -np_{i0}p_{j0}\text{\hspace *{1.3cm}   for  } i \neq j
      \end{cases}$$
where
\begin{equation}
 r_i=\mu_i/ \alpha_{2i}
 \end{equation}
  is the ratio of the first moment of the distribution of weights of events  $\mu_i$ to the the second moment $\alpha_{2i}$ for a particular bin $i$.
Notice that for usual histograms the ratio of moments $r_i$ is equal
to 1 and the covariance matrix coincides with the covariance matrix
of the multinomial distribution.

The multivariate  statistic is represented as
\begin{equation}
(\textbf{W}-n\textbf{p}_0)^t\mathbf{\Gamma_k^{-1}}(\textbf{W}-n\textbf{p}_0),
\end{equation}
where \\ $\textbf{W}=(W_1,\ldots,W_{k-1},W_{k+1},\ldots,W_{m})^t$,
$\textbf{p}_0=(p_{10},\ldots,p_{k-1,0},p_{k+1,0},\ldots,p_{m0})^t$
and $\mathbf{\Gamma_k}=(\gamma_{ij})_{(m-1)\times(m-1)}$ is the covariance
matrix for a histogram without bin $k$. The matrix $\mathbf{\Gamma_k}$ has the form
\begin{equation}
\mathbf{\Gamma_k}=n\,\textrm{diag}\,(\frac{p_{10}}{r_1}, \ldots
,\frac{p_{k-1,0}}{r_{k-1}},\frac{p_{k+1,0}}{r_{k+1}}, \ldots ,
\frac{p_{m 0}}{r_m})-n \textbf{p}_0\textbf{p}_0^t.
\end{equation}
 The special form of this matrix permits one  to find analytically the inverse matrix
\begin{equation}
\mathbf{\Gamma_k^{-1}}=\frac{1}{n}\textrm{diag}\,(\frac{r_1}{p_{10}}, \ldots
,\frac{r_{k-1}}{p_{k-1,0}},\frac{r_{k+1}}{p_{k+1,0}}, \ldots ,\frac{r_m}{p_{m 0}})+\frac{1}{n(1-\sum_{i \neq k}r_i
p_{i0})}\mathbf{r}\mathbf{r}^t,
\end{equation}
 where $\textbf{r}=(r_1,\ldots,r_{k-1},r_{k+1},\ldots,r_{m})^t$.

 After that, the multivariate statistic can be written as
\begin{equation}
X^2_k= \sum_{i \neq k} r_i \frac{(W_i-np_{i0})^2}{np_{i0}}+\frac{(\sum_{i \neq k} r_i(W_i-np_{i0}))^2}
{n(1-\sum_{i \neq k}r_ip_{i0})},\label{stdd1}
\end{equation}
and   can also be transformed to form
\begin{equation}
 X^2_k=\frac{1}{n} \sum_{i \neq k} \frac{r_iW_i^2}{p_{i0}}+\frac{1}{n}
\frac{(n-\sum_{i \neq k}r_iW_i)^2}{1-\sum_{i \neq k}r_i
p_{i0}}-n\label{stdd2}
\end{equation}
which is convenient for numerical calculations. Asymptotically the
vector $\textbf{W}$ has a normal distribution $\mathcal{N}(n
\textbf{p}_0,\mathbf{\Gamma_k^{1/2}})$   \cite{robins} and therefore the test
statistic (\ref{stdd1}) has $\chi^2_{m-1}$ distribution if hypothesis
${H_0}$ is true
\begin{equation}
 X_k^2 \sim \chi^2_{m-1}.
\end{equation}
For usual histograms when $r_i=1$,
$i=1,\ldots, m$ the statistic (\ref{stdd1}) is Pearson's chi-square
statistic (\ref{basic11}).

The expectation value of statistic (\ref{stdd1}), as shown in Ref. \cite{gagunash},  is equal to
 \begin{equation}
\textrm E[X^2_k]=m-1,
 \end{equation}
 as for Pearson's test \cite{kendall}.

The ratio of moments $r_i=\mu_i/ \alpha_{2i}$, that is used for the test statistic calculation, is not known in majority of cases. An estimation of $r_i$ can be used:
\begin{equation}
\hat{r_i}=W_i/W_{2i},
\end{equation}
where $W_{2i}=\sum_{k=1}^{n_i}w_i^2(k)$.

Let us now replace $r_i$ with the estimate $\hat
r_i$ and denote the estimator of matrix $\mathbf{\Gamma_k}$ as
${\mathbf{\hat \Gamma_k}}$. Then for positive definite matrices $
{\mathbf{ \hat\Gamma_k}}$, $k=1,\ldots,m$ the test statistic is given as
 \begin{equation}
\hat{X^2_k}= \sum_{i \neq k} \hat r_i \frac{(W_i-np_{i0})^2}{np_{i0}}+\frac{(\sum_{i \neq k}
 \hat r_i(W_i-np_{i0}))^2} {n(1-\sum_{i \neq k} \hat r_i p_{i0})}.\label{stdd333}
\end{equation}

Formula (\ref{stdd333}) for usual histograms does not depend on the
choice of the excluded  bin, but for weighted histograms there can
be a dependence. A test statistic that is invariant to the choice of
the excluded bin and at the same time is a Pearson's chi square
statistic (\ref{basic11})   for the unweighted histograms can be represented as the median value for the set of statistics $\hat X^2_k$ (\ref{stdd333}) with positive definite matrixes ${\mathbf{\hat\Gamma_k}}$
\begin{equation}
\hat X_{Med}^2= \textrm {Med }\, \{\hat X_1^2,  \hat X_2^2,  \ldots , \hat
X_m^2\}.\label{stdav}
\end{equation}
Statistic $\hat X_{Med}^2$ first time was proposed in Ref. \cite{gagunash} and approximately has $\chi^2_{m-1}$ distribution if hypothesis ${H_0}$ is true
\begin{equation}
\hat X_{Med}^2 \sim \chi^2_{m-1}.
\end{equation}
The usage of $\hat X_{Med}^2$ to test the hypothesis $H_0$ with a given
significance level is equivalent to making a decision by voting. It was noticed that size of test can be slightly  greater than nominal value of size of test even for large value of total number of events $n$.

\subsection{New generalizations of Pearson's chi-square test for weighted histograms}
  Set of statistics $\{\hat X_1^2,  \hat X_2^2,  \ldots , \hat X_m^2 \}$,  with positive definite matrixes ${\mathbf{\hat \Gamma_k}}$ only, is used for calculating the median statistic $\hat X_{Med}^2$ (\ref{stdav}).
   It can be used  for  any weighted histograms, including histograms with unweighted entries. One bin is excluded because the full covariance matrix of an unweighted histogram is singular and hence can not  be inverted.

 Let us consider estimation of a full covariance matrix $\mathbf{\hat \Gamma}$  for the weighted histogram with more detail. The symmetric matrix is positive definite if the minimal eigenvalue of the matrix larger then 0.  We denote minimal eigenvalue of the matrix $n^{-1}\mathbf{\hat \Gamma}$  by $\lambda_{min}$ then follow to  Ref. \cite{wilkin} it can be shown that
 \begin{equation}
 \min_i\{\frac{p_{i0}}{\hat r_{i}}\} - \sum_{i=1}^{m}p_{i0}^2 \leq \lambda_{min} \leq \min_i\{ \frac{p_{i0}}{\hat r_{i}}\}.
 \end{equation}
 and the eigenvalue $\lambda_{min}$ is the root of  secular equation
 \begin{equation}
  1-\sum_{i=1}^m \frac{p_{i0}^2}{p_{i0}/\hat r_i-\lambda}=0. \label{sekular}
 \end{equation}

 In case of a histogram with unweighted entries, all $\hat r_i=1$ and $\lambda=0$ is zero of equation (\ref{sekular}). Matrix $\mathbf{\hat \Gamma}$ for this case is not positive definite and is singular, but matrix $\mathbf{\hat \Gamma_k}$  is positive definite and therefore invertible. Number of events $n_i$ in bins  of usual histogram satisfy to  equation
 $n_1+n_2,...,+n_m=n$ that is why the covariance matrix of multinomial  distribution is not positive definite and is singular.

  Matrix $\mathbf{\hat \Gamma}$ for a histogram with weighted entries can be also non-positive definite. There are two reasons why this can be. First of all, the  total sums of weights $W_i$ in bins  of a weighted histogram  are related with each other, because satisfy the equation $\textrm E[\sum_{i=1}^m{W_i}]=n$ and  second, due fluctuations of  matrix elements.

   The test statistic obtained with full matrix $\mathbf{\hat \Gamma}$ is unstable and can have large variance especially for the case of low number $n$ of events in a histogram.

   The fact the matrix is not positive definite is equivalent to the fact that the minimal eigenvalue $\lambda_{min}$  of  the matrix $\mathbf{\hat \Gamma}$ is  $\leq 0$. A case when the minimal  eigenvalue is  positive but rather  small is also not desirable, especially for computer calculations.\\

 Due to the above mentioned reasons it is wise to use the test statistic  for a weighted histograms
\begin{equation}
 \hat X^2 =\hat X^2_k =
   \mathlarger{\sum}\limits_{i \neq k} \hat r_i \dfrac{(W_i-np_{i0})^2}{np_{i0}}+\dfrac{(\sum_{i \neq k}
 \hat r_i(W_i-np_{i0}))^2} {n(1-\sum_{i \neq k} \hat r_i p_{i0})}\label{newx}
\end{equation}
for $k$ where
\begin{equation}
\frac{p_{k0}}{\hat r_k}=\min_i\{\frac{p_{i0}}{\hat r_{i}}\}.
\end{equation}
A secular equation for the new minimal eigenvalue can be solved numerically, by bisection method,   to check whether a matrix ${\mathbf{\hat \Gamma_k}}$ is ​​positive definite or  not. Numerical experiments show that it is very rare that the matrix ${\mathbf{\hat \Gamma_k}}$ is not positive definite and it happens only  for histograms with a small number $n$ of events in a histogram.
If hypothesis $H_0$ is valid, statistic  $\hat X^2$ asymptotically has distribution
\begin{equation}
 \hat X^2 \sim
    \chi_{m-1}^2.
\end{equation}

It is plausible that power of the new test is not lower than  power of tests with statistic $\hat X_{Med}^2$  and  with other statistics $\{\hat X^2_i, i \neq k\} $. The distribution of the statistic $\hat X^2$ is closer to $\chi_{m-1}^2$  then  distribution of median statistic $\hat X_{Med}^2$.  Also the statistic $\hat X^2$  is easier to calculate than  the statistic $\hat X_{Med}^2$.

\section{Goodness of fit tests for weighted histograms with deviations from main model}

Here, different deviations from the main model of weighted histograms will be considered as well as goodness of fit tests for those cases.

\subsection{Goodness of fit test for weighted histogram with unknown normalization}

In practice one is often faced with the case that all weights of events are  defined up to an unknown normalization constant $C$ see Ref. \cite{gagunash}.
It happens because in some cases of computer simulation  is rather difficult give analytical formula for the PDF, but the PDF up to multiplicative constant is possible, that is enough for the generation of events according to the PDF, for example, by very popular Neumann's method \cite{neiman}.
 For the goodness of fit test it  means    that  if hypothesis $H_0$ is valid
\begin{equation}
\textrm E [W_i]\cdot C=np_{i0}, \; i=1,\ldots,m. \label{ew}
\end{equation}
with unknown constant $C$.
 Then the test statistic (\ref{stdd2}) can be
written as
\begin{equation}
_c\hat{X^2_k}= \sum_{i \neq k} \hat r_i \frac{(W_i-np_{i0}/C)^2}{np_{i0}/C}+\frac{(\sum_{i \neq k}
 \hat r_i(W_i-np_{i0}/C))^2} {n(1-C^{-1}\sum_{i \neq k} \hat r_i p_{i0})}.\label{stdd33}
\end{equation}

An estimator for the constant $C$ can be
found by minimizing Eq. (\ref{stdd33}).
\begin{equation}
\hat C_k=\sum_{i \neq k}\hat{r}_ip_{i0}+ \sqrt{\frac{\sum_{i
\neq k}\hat{r}_ip_{i0}}{\sum_{i \neq k}\hat{r}_i W_i^2/p_{i0}}}(n-\sum_{i \neq
k}\hat{r}_i W_i), \label{consti}
\end{equation}
where $\hat C_k$ is an estimator of $C$.  Substituting (\ref{consti}) to  (\ref{stdd33}), we get the test statistic
\begin{equation}
_c\hat{X^2_k}= \sum_{i \neq k} \hat r_i \frac{(W_i-np_{i0}/\hat C_k)^2}{np_{i0}/\hat C_k}+\frac{(\sum_{i \neq k}
 \hat r_i(W_i-np_{i0}/\hat C_k))^2} {n(1-\hat C_k^{-1}\sum_{i \neq k} \hat r_i p_{i0})}.\label{stdd}
\end{equation}
The statistic (\ref {stdd}) has a $\chi^2_{m-2}$ distribution if hypothesis ${H_0}$ is valid.

Formula  (\ref{stdd}) can be also transformed to
\begin{equation}
_c\hat{X}^2_k =\frac{s_k^2}{n}+2s_k, \label{sss}
\end{equation}
where
\begin{equation}
s_k=\sqrt{\sum_{i \neq k}\hat{r}_i p_{i0} \sum_{i \neq k}
\hat{r}_i{W}_i^2/p_{i0}} - \sum_{i \neq
 k}\hat{r}_i {W}_i
\end{equation}
which is convenient for calculations, see \cite{gagunash}.
Median statistics can be used for the same reason as in section 2.2
\begin{equation}
 _c\hat{X}_{Med}^2= \textrm {Med }\, \{_c\hat X_1^2,\,  _c\hat X_2^2,  \ldots ,  \,_c\hat X_m^2\}. \label{stdav2}
\end{equation}
and has approximately $\chi^2_{m-2}$ distribution if hypothesis ${H_0}$ valid, see Ref. \cite{gagunash}
\begin{equation}
_c\hat{X}_{Med}^2  \sim  \chi_{m-2}^2.
\end{equation}
\subsection{New goodness of fit test for weighted histogram with unknown normalization}

The new estimator of constant $C$ is
\begin{equation}
\hat C=\sum_{i \neq k}\hat{r}_ip_{i0}+ \sqrt{\frac{\sum_{i
\neq k}\hat{r}_ip_{i0}}{\sum_{i \neq k}\hat{r}_i W_i^2/p_{i0}}}(n-\sum_{i \neq
k}\hat{r}_i W_i),
\end{equation}
for $k$ where
\begin{equation}
\frac{p_{k0}}{\hat r_k}=\min_i\{\frac{p_{i0}}{\hat r_{i}}\}.
\end{equation}

 And  the test statistic  can be written as
 \begin{equation}
 _c\hat{X^2} =
 {\mathlarger{\sum}\limits_{i \neq k}} \hat r_i \dfrac{(W_i-np_{i0}/\hat C)^2}{np_{i0}/\hat C}+\dfrac{(\sum_1^m
 \hat r_i(W_i-np_{i0}/\hat C))^2} {n(1- \hat C^{-1}\sum_1^m \hat r_i. p_{i0})}\label{unknown}
\end{equation}

Statistic  $_c\hat X^2$ asymptotically has $\chi^2_{m-2}$ distribution if hypothesis $H_0$ is valid
 \begin{equation}
 _c\hat{X}^2 \sim \chi^2_{m-2}.
\end{equation}

\subsection{Goodness of fit test for weighted Poisson histograms}

Poisson  histogram \cite{cousine} can be defined as histogram with  multi-Poisson distributions of a number of events for bins
\begin{equation}
P(n_1,\ldots ,n_m)=\prod_{i=1}^m e^{-n_0p_i}(n_0p_i)^{n_i}/n_i!,
\end{equation}
where $n_0$ is a free parameter.
The discrete probability distribution function (probability mass function)  of a Poisson histogram can be represented as a product of two probability functions: a Poisson probability mass function  for a number of events $n$  with parameter  $n_0$ and a multinomial probability mass function of the number of events for bins of the  histogram, with total number of events equal to $n$, see Ref. \cite{kendall}
\begin{equation}
P(n_1,\ldots ,n_m)=e^{-n_0}(n_0)^{n}/n!\times\frac{n!}{n_1!n_2! \ldots n_m!} \; p_1^{n_1}\ldots p_m^{n_m}.
\end{equation}
 A Poisson histogram can be obtained as a result of two random experiments, namely, where the first experiment with Poisson probability mass function gives us the total number of events in histogram $n$,   and then a histogram is obtained as a result of a random experiment with PDF $p(x)$ and the total number of events is equal to $n$.

As in  the case of multinomial histograms, also for Poisson histograms there is  the problem of goodness of fit test with the hypothesis:
\begin{equation}
H_0: p_1=p_{10},\ldots, p_{m-1}=p_{m-1,0} \text{  vs.  } H_a: p_i \neq p_{i0} \text{  for some  } i,
\end{equation}
where $p_{i0}$ are specified probabilities, and $\sum_{i=1}^{m}p_{i0}=1$.
If $n_0$ is known, then the statistic, see Ref. \cite {zech}:
\begin{equation}
X_{pois}^2=\sum_{i=1}^{m} \frac{(n_i-n_0p_{i0})^2}{n_0p_{i0}}, \label{basic}
\end{equation}
can be used and has asymptotically  a $\chi^2_{m}$ distribution if the hypothesis $H_0$ is valid
\begin{equation}
X_{pois}^2  \sim \chi^2_{m}.
\end{equation}

 The hypothesis $H_0$ becomes complex if parameter $n_0$ is unknown for the Poisson histogram.  This is an opposite situation to the  case of a multinomial histogram,  where the hypothesis is simple.

In \cite{zech} there are proposed statistics for goodness of fit test for a weighted Poisson  histogram with known parameter $n_0$
\begin{equation}
 X^2_{corr0}=  \sum_{i=1}^m \frac{(W_i-n_0p_{i0})^2}{W_{2i}n_0p_{i0}/W_i}, \label{stx15}
\end{equation}
and for the case the $n_0$ is not known:
\begin{equation}
 X^2_{corr}=  \sum_{i=1}^m \frac{(W_i-\hat n_0p_{i0})^2}{W_{2i}\hat n_0p_{i0}/W_i}, \label{stx12}
\end{equation}
with estimation of $n_0$ obtained by minimization of equation (\ref{stx15})
\begin{equation}
 \hat n_0= \left[\frac{\sum_{i=1}^m W_i^3/(W_{2i}p_{0i})}{\sum_{i=1}^m W_ip_{0i}/W_{2i}}\right]^{1/2}. \label{sx}
\end{equation}
The distribution of statistic $X^2_{corr0}$  in case hypothesis $H_0$ is valid
\begin{equation}
X_{corr0}^2  \sim \chi^2_{m}
\end{equation}
and for the statistic $X^2_{corr}$ is
\begin{equation}
X_{corr}^2  \sim \chi^2_{m-1}
\end{equation}
according Ref. \cite{zech}.

    Generally, the power of the tests for Poisson histograms will be slightly lower than for multinomial histograms with the number of events $n=n_0$ which is explained by the fact that  the total number of events for Poisson histograms fluctuates.

The choice of the type of the histogram depends on what type of a physical experiment is produced. If the number of events $n$ is constant, then it is a multinomial histogram; if the number of events $n$  is a random value that has  Poisson distribution, then it is a Poisson histogram.

 A weighted histogram very often  is the result of modeling and the number of simulated events is known exactly, and therefore the choice of a multinomial histogram is reasonable. It is also reasonable to use tests developed for the multinomial histograms in the case, if the number of events $n$ is random value but with unknown  distribution \cite{edie}.

\section{Restriction for goodness of fit tests applications}

For the histograms with unweighted entries, the use of Pearson's chi-square test (\ref{basic11}) is inappropriate if any expected frequency $np_{i0}$ is below 1 or if the expected frequency is less than 5 in more than 20\% of bins \cite{moore}.

Restrictions for weighted histograms, due to fluctuation of the estimation of ratio of moments $\hat r_i$,  can be made stronger. Namely, the use of  new chi-square tests (\ref{newx}) and (\ref{unknown}) is inappropriate if any expected frequency  $\textrm E[n_i]$ is less than 5.

Following Ref. \cite{cochran} a disturbance is regarded as unimportant when the nominal size of the test is 5\% and the size of the test  lies between 4\% and 6\% for a goodness of fit tests.
\section{Numerical evaluation of the tests' power and sizes}

The main parameters which characterizes the  effectiveness of a test are size and  power.

The nominal significance level was taken to be equal to 5\% for calculating of size of tests in presented numerical examples.
Hypothesis $H_0$ is rejected if test statistic $\hat X^2$ is larger
than some threshold. Threshold $k_{0.05 }$ for a given nominal
size of test 5\% can be  defined from the equation
\begin{equation}
0.05 = P\,(\chi^2_l>k_{0.05})=\int_{k_{0.05}}^{+\infty}
\frac{x^{l/2-1} e^{-x/2}}{2^{l/2}\Gamma(l/2)}dx, \label {kalfa}
\end{equation}
where $l=m-1$.

 Let us define the test size $\alpha$ for a given nominal test
size 5\% as the probability
\begin{equation}
\alpha =  P\,(\hat X^2>k_{0.05}|H_0).\label {alfas}
\end{equation}
This is the probability that hypothesis $H_0$ will be rejected if
 the distribution of weights $W_i$ for bins of the histogram  satisfies hypothesis $H_0$.
Deviation of the test size from the nominal test size is an
important test characteristic.

A second important test characteristic is the power. Let us define the test power  as
\begin{equation}
P\,(\hat X^2 >k_{0.05}|H_a). \label {beta}
\end{equation}
This is the probability that  hypothesis $H_0$ will be rejected if
the distribution of weights $W_i$ for bins of the  histogram does
not satisfy hypothesis $H_0$.

Notice that the power calculated by formula (\ref{beta}) can give misleading result in case of comparing of different tests.
  To overcome this problem here we define the power of test $\pi$ as
\begin{equation}
\pi = P\,(\hat X^2 >\mathcal{K}_{0.05}|H_a) \label {beta2}
\end{equation}
with the threshold $\mathcal{K}_{0.05}$ calculated by Monte-Carlo method
from equation
\begin{equation}
0.05 =  P\,(\hat X^2>\mathcal{K}_{0.05}|H_0).\label {alfass}
\end{equation}
 All definitions proposed  above for statistics $\hat X^2$  can be used for other test statistics with appropriate number of degree of freedom $l$ in the formula (\ref{kalfa}).

The size and  power of tests depend on the number of events and the binning that was discussed for usual histograms in Ref. \cite{kendall}. The power for weighted histograms also depends on  the choice of PDF  $g(x)$ (subsection 1.1) or $g_{tr}$ (subsection 1.2)  and can be even higher than for histograms with unweighted entries as well as lower.
Below we demonstrate  two  examples of an application of the previously discussed tests. The  size and power of the tests are calculated for a different total number of events in the histograms.
In numerical examples were demonstrated applications of:
\begin{itemize}
\item Pearson's goodness of fit test \cite{pearson}, see subsection 2.1 and first paragraph of section 2. The test statistic is $X_2$ (\ref{basic11}).

\item goodness of fit test for weighted histograms with normalized weights \cite{gagunash}, see subsection 2.2. The test statistic is $\hat X_{Med}^2$ (\ref{stdav}).

\item goodness of fit test  for weighted histograms  with unnormalized weights  \cite{gagunash}, see subsection 3.1. The test statistic is $_{c}\hat X_{Med}^2$ (\ref{stdav2}).

\item new goodness of fit test for weighted histograms with normalized weights, see subsection 2.3. The test statistic is $\hat X^2$(\ref{newx}).
 \item new goodness of fit test for weighted histograms  with unnormalized weights, see subsection 3.2. The test statistic is $_{c}\hat X^2$(\ref{unknown}).

\item goodness of fit test  for Poisson histograms with unweighted entries and known parameter $n_0$ \cite{zech},  see subsection 3.3. The test statistic is $X_{pois}^2$(\ref{basic}).
 \item goodness of fit test for weighted Poisson histograms with known parameter $n_0$ \cite{zech},  see subsection 3.3. The test statistic is $X_{corr0}^2$(\ref{stx15}).
 \item  goodness of fit test  for weighted Poisson histograms with unknown parameter $n_0$ \cite{zech},  see subsection 3.3. The test statistic is $X_{corr}^2$(\ref{stx12}).
\end{itemize}
The published program, see Ref. (\cite{gagunash3}), was  used for the calculation of the test statistics with minor modification  needed for the new tests.

\subsection{Numerical example 1}
\begin{figure}
\vspace *{-3 cm}
\centering
\includegraphics[width=1. \textwidth]{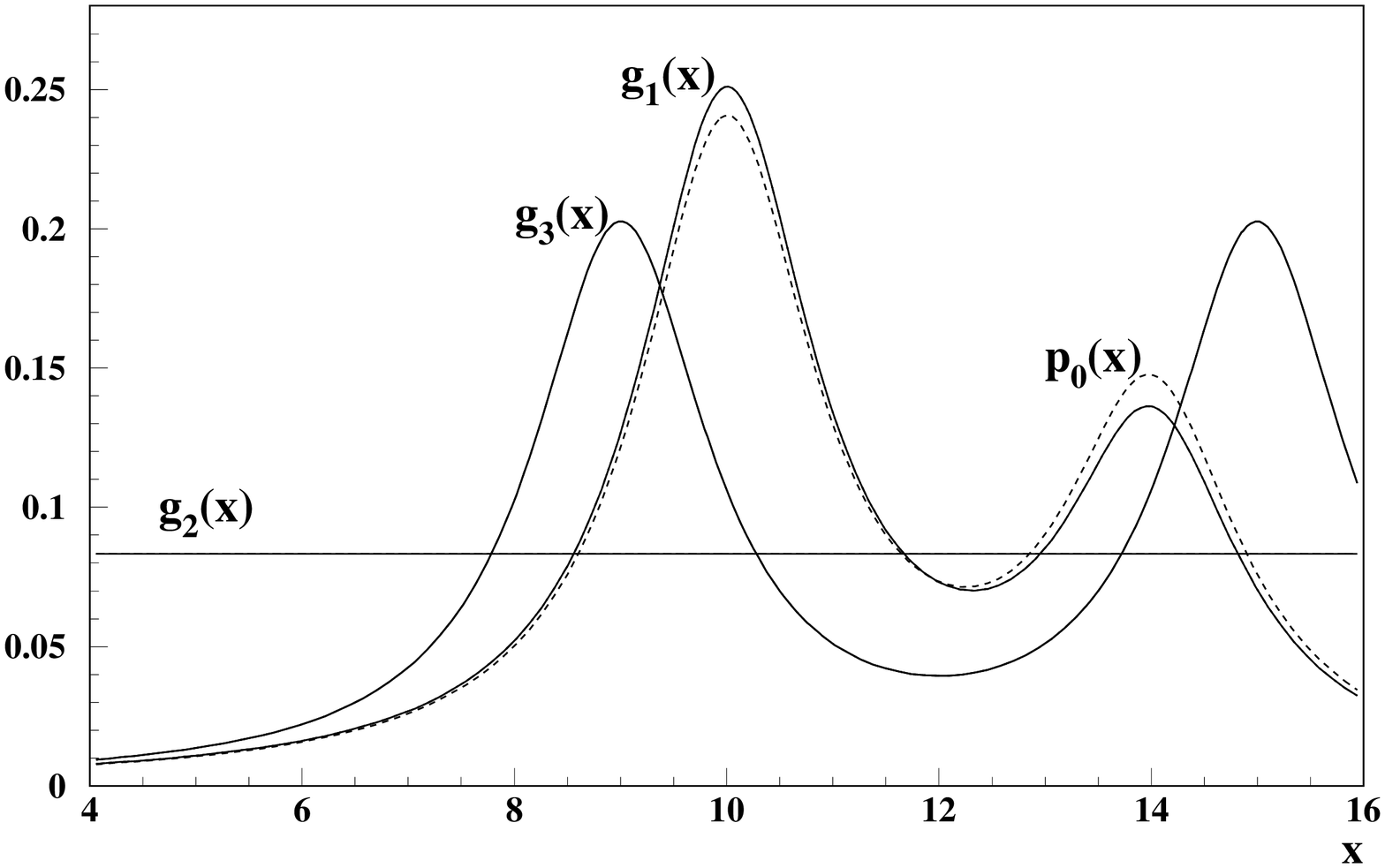}
\vspace *{-1.9cm}
 \caption {Probability density functions $g_1(x)=p(x)$, $g_2(x)$,
 $g_3(x)$ used for events generation and PDF $p_0(x)$ for hypothesis $H_0$ (dashed line)}
 \label{fig:picture1}
\end{figure}

\begin{figure}
\centering
\vspace *{-3.5cm}
 \includegraphics[width=1. \textwidth]{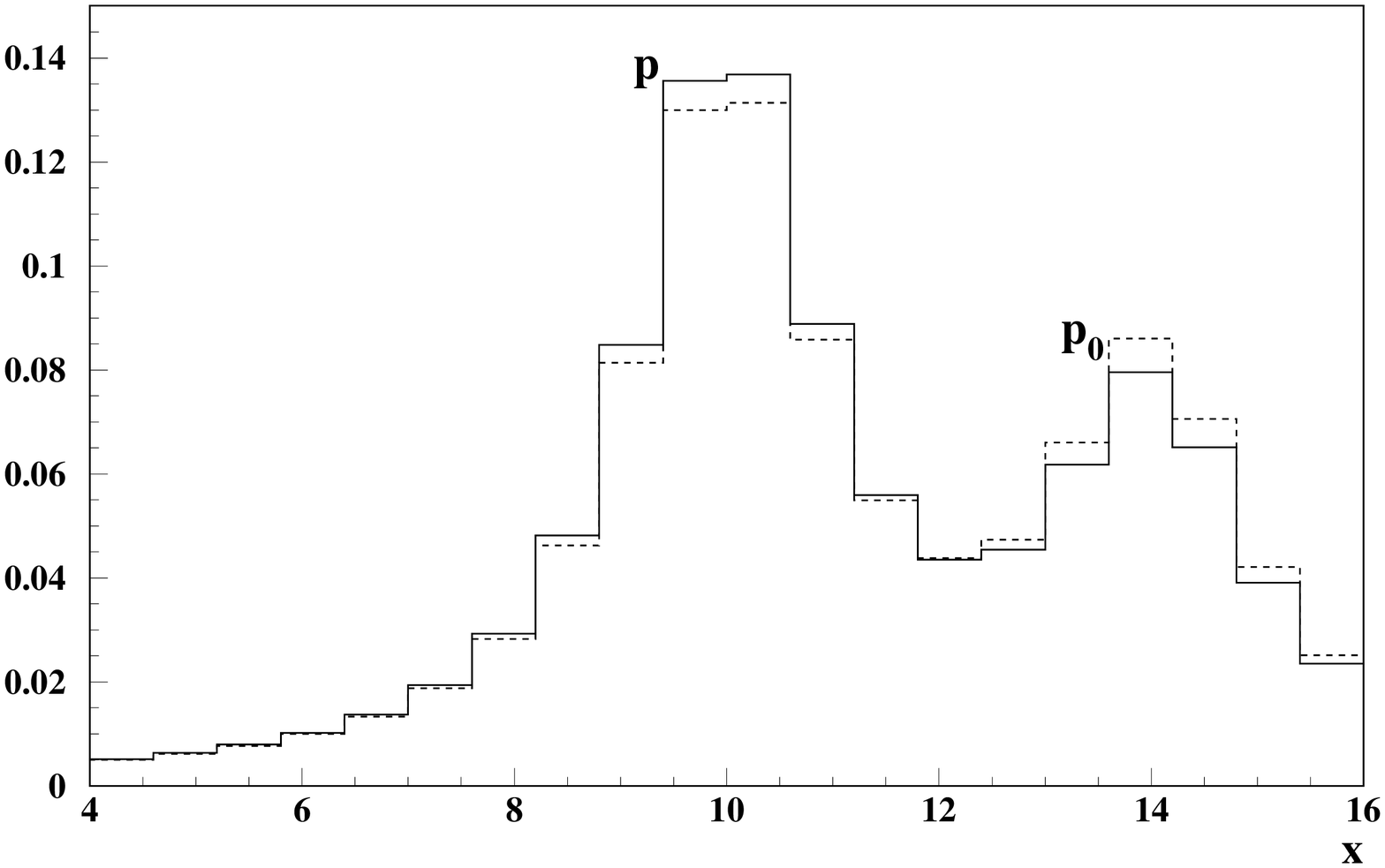}
\caption {Probabilities $p_i, i=1,...,20$ for the PDF $p(x)$
(solid line)  and $p_{0i}, i=1,...,20$ for the PDF $p_{0}(x)$ (dashed line)}
\label{fig:hyalbw}
\end{figure}

A simulation study was done for the example from Ref. \cite{gagunash}.
 Weighted histograms described in subsection 1.1, are used here.
The PDF for hypothesis $H_0$ is:
\begin{equation}
p_0(x) \propto \frac{2}{(x-10)^2+1}+\frac{1.15}{(x-14)^2+1}\label{basicpar}
\end{equation}
against alternative $H_a$:
\begin{equation}
p(x) \propto \frac{2}{(x-10)^2+1}+\frac{1}{(x-14)^2+1}
\end{equation}
represented by the weighted histogram.  Both  PDF's are defined on the interval $[4,16]$.
 A calculation was done for three cases of a PDF, used for the event generation, see Fig.~\ref{fig:picture1}
 \begin{equation}
 g_1(x)=p(x) \label{case1}
 \end{equation}
 \begin{equation}
 g_2(x)=1/12 \label{case2}
 \end{equation}
\begin{equation}
g_3(x) \propto \frac{2}{(x-9)^2+1}+\frac{2}{(x-15)^2+1} \label{case3}.
\end{equation}
Distribution (\ref{case1}) gives an unweighted histogram. Distribution (\ref{case2}) is a uniform distribution on the interval $[4, 16]$. Distribution (\ref{case3}) has  the same type of parameterizations as Eq. (\ref{basicpar}), but with different values of the parameters.
Histograms with 20 bins  and  equidistant binning were used. At Fig.~\ref{fig:hyalbw}  presented probabilities  $p_i, i=1,...,20$ for the PDF $p(x)$ and $p_{0i}, i=1,...,20$ for the PDF $p_{0}(x)$. Size and power of tests with statistics $\hat X^2$ (\ref{newx}),   $_{c}\hat X^2$ (\ref{unknown}),  $\hat X_{Med}^2$ (\ref{stdav}) and  $_{c}\hat X_{Med}^2$ (\ref{stdav2}) were  calculated  for  weighted histograms with weights of events equal to $p(x)/g_2(x)$ and $p(x)/g_3(x)$.
 Statistics $\hat X^2$ (\ref{newx}) and $\hat X_{Med}^2$ (\ref{stdav}) coincide  with Pearson's statistic $X^2$ (\ref{basic11}) and were used for  histograms with unweighted entries.
The results of calculations  for 100000 runs are presented in Table~\ref{tab:result31}. \\
 \begin{table}[h]
\centering
\caption{Numerical example 1. Size ($\alpha$) and power ($\pi$) of different test statistics $X^2$(\ref{basic11}), $\hat X^2$(\ref{newx}), $_{c}\hat X^2$(\ref{unknown}), $\hat X_{Med}^2$(\ref{stdav}), $_{c}\hat X_{Med}^2$(\ref{stdav2}) obtained for different weighted functions $w(x)$. Italic type marks a size of test  with inappropriate number of events in the bins of  histograms.}
\label{tab:result31}
\vspace *{0.4 cm}
\scriptsize
\begin{tabular}{c|c|l|rrrrrrrrr|c}
\textnumero & & $n$ &   200 &   400 &   600 &   800 &  1000 &  3000 &  5000 &  7000 &  9000 & w(x) \\
 \hline
{\multirow{2}{*}{$1$}} & {\multirow{2}{*} {$X^2$}} & $\alpha$ & \textit{5.7} &   \textit{5.4} &   5.3 &   5.2 &   5.2 &   5.1 &   5.0 &   5.0 &   5.1 & {\multirow{2}{*}{$1$}} \\
& & $\pi$ &  \textit{6.0} &  \textit{7.1} &   8.2 &   9.8 &  11.2 &  29.9 &  52.7 &  71.6 &  84.9 & \\

 \hline
 \hline
 {\multirow{2}{*}{$2$}} & {\multirow{2}{*} {$\hat X^2$}} & $\alpha$ &  5.5 &   5.3 &   5.2 &   5.1 &   5.0 &   5.1 &   5.1 &   5.1 &   4.9 &{\multirow{8}{*}{$\frac{p(x)}{g_2(x)}$}} \\
 &&$\pi$ &   6.1 &   7.0 &   8.2 &   9.2 &  10.5 &  26.2 &  45.8 &  64.0 &  78.7 &\\
 \cline{1-12}
 {\multirow{2}{*}{$3$}} & {\multirow{2}{*} {$_{c}\hat X^2$}} & $\alpha$                         &   5.0 &   5.1 &   5.0 &   5.0 &   4.9 &   5.0 &   5.0 &   5.2 &   4.9 &\\
 &&$\pi$ &   6.0 &   7.0 &   8.1 &   9.1 &  10.4 &  26.0 &  45.6 &  63.0 &  78.1 &\\

 \cline{1-12}
 {\multirow{2}{*}{$4$}} & {\multirow{2}{*} {$\hat X_{Med}^2 $}} & $\alpha$                                   &   5.4 &   5.4 &   5.3 &   5.2 &   5.1 &   5.3 &   5.2 &   5.3 &   5.0 &\\
 &&$\pi$ &   6.0 &   6.9 &   8.0 &   9.1 &  10.3 &  25.7 &  45.3 &  63.1 &  78.2 &\\
 \cline{1-12}
 {\multirow{2}{*}{$5$}} & {\multirow{2}{*} {$ _{c}\hat X_{Med}^2$}} & $\alpha$                                   & 5.6 &   5.8 &   5.7 &   5.7 &   5.5 &   5.7 &   5.7 &   5.8 &   5.5 &\\
 &&$\pi$ &   5.9 &   6.9 &   8.0 &   9.1 &  10.2 &  25.4 &  44.9 &  62.5 &  77.5 &\\

 \hline
 \hline
  {\multirow{2}{*}{$6$}} & {\multirow{2}{*} {$ \hat X^2$}} & $\alpha$                                             &  \textit{ 7.3} & \textit{  6.6} & \textit{  6.1 }&   5.8 &   5.6 &   5.2 &   5.2 &   4.9 &   5.0 & \normalsize{\multirow{8}{*}{$\frac{p(x)}{g_3(x)}$}} \\
 &&$\pi$ & \textit{16.2} & \textit{29.7} &  \textit{40.1} &  48.5 &  56.1 &  95.7 &  99.8 & 100.0 & 100.0 &\\
 \cline{1-12}
 {\multirow{2}{*}{$7$}} & {\multirow{2}{*} {$_{c}\hat X^2 $}} & $\alpha$                                               &  \textit{ 4.7} &  \textit{ 4.9} & \textit{  5.0} &   5.0 &   5.1 &   5.1 &   5.0 &   4.9 &   5.0 &\\
 &&$\pi$ &   \textit{6.9} &   \textit{8.3} &   \textit{9.9} &  11.6 &  13.4 &  36.5 &  61.6 &  80.5 &  91.2 &\\
 \cline{1-12}
 {\multirow{2}{*}{$8$}} & {\multirow{2}{*} {$\hat X_{Med}^2 $}} & $\alpha$                                                 & \textit{  5.5} &  \textit{ 5.3} & \textit{  5.4} &   5.3 &   5.5 &   5.4 &   5.2 &   5.3 &   5.2 &\\
  &&$\pi$ &   \textit{7.9} &  \textit{11.8} &  \textit{15.8} &  20.2 &  25.0 &  75.3 &  96.6 &  99.8 & 100.0 &\\
 \cline{1-12}
 {\multirow{2}{*}{$9$}} & {\multirow{2}{*} {$_{c}\hat X_{Med}^2$}} & $\alpha$                                                  &  \textit{ 5.4} &  \textit{ 5.5} &  \textit{ 5.7} &   5.6 &   5.8 &   5.7 &   5.6 &   5.7 &   5.5 &\\
 &&$\pi$ &   \textit{6.8} &   \textit{8.4} &   \textit{9.7} &  11.4 &  13.1 &  36.0 &  60.7 &  79.2 &  90.7 &
\end{tabular}
\end{table}
Conclusion and  interpretation of results presented in Table \ref{tab:result31}.

\begin{itemize}
\item The size of new tests $\hat X^2$ (\ref{newx})(rows 2, 6) and $ _{c}\hat X^2$(\ref{unknown})(row 3, 7) are generally closer to nominal value 5\% then median tests $\hat X_{Med}^2$ (\ref{stdav})(rows 4, 8) and $_{c}\hat X_{Med}^2$ (\ref{stdav2})(rows 5, 9) when the application of the test  satisfies  restrictions formulated in section 4.
\item The power of new tests $\hat X^2$ (\ref{newx}) (rows 2, 6) are greater than for analogous median tests  $\hat X_{Med}^2$ (\ref{stdav})(rows 4, 8). The  power of tests $ _{c}\hat X^2 $(\ref{unknown})(rows 3, 7) are greater than for analogous  median tests $_{c}\hat X_{Med}^2$ (\ref{stdav2})(rows 5, 9).
\item The power of all tests calculated for histograms with weights of events equal to $ p(x)/g_2(x)$ (rows 2-5) are lower then for histogram with unweighted entries (row 1), but the power of all tests calculated for histograms with weights of events
equal to $p(x)/g_3(x)$ (rows 6-9) are greater. The explanation is that in latter case  we  increase the statistics of events for domains with high deviation of the distribution presented by the histogram from the tested  distribution.
\end{itemize}

Properties of tests  in applications  to Poisson histograms  with the same weighted functions and distributions of events were investigated. In this case, the total number of events $n$ is random  and was simulated according Poisson distribution for a given parameter $n_0$. Size and power of tests $X_{pois}^2$ (\ref{basic}), $X_{corr0}^2$ (\ref{stx15}) with exactly known parameter $n_0$ and  $X_{corr}^2$ (\ref{stx12})  developed ad hoc for the Poisson histogram in \cite{zech} also was calculated.
Results of the calculations are presented in Table~\ref{tab:result25} .
 \begin{table}[h]
\centering
\caption{Numerical example 1. Size ($\alpha$) and power ($\pi$) of different test statistics $X_{pois}^2$(\ref{basic}), $X^2$(\ref{basic11})  , $X_{corr0}^2$(\ref{stx15}), $X_{corr}^2$(\ref{stx12}) , $\hat X^2$(\ref{newx})  , $_{c}\hat X^2 $(\ref{unknown})  in  application for Poisson histograms. Italic type marks a size of test  with inappropriate number of events in the bins of  histograms.}
\label{tab:result25}
\vspace *{0.4 cm}
\scriptsize
\begin{tabular}{c|c|l|rrrrrrrrr|c}
\textnumero & & $n_0$ &   200 &   400 &   600 &   800 &  1000 &  3000 &  5000 &  7000 &  9000 & w(x) \\
 \hline
 {\multirow{2}{*}{$1$}} & {\multirow{2}{*} {$X_{pois}^2$}}& $\alpha$  &   6.0 &   5.6 &   5.2 &   5.2 &   5.2 &   5.1 &   5.1 &   5.1 &   5.1 &{\multirow{4}{*}{$1$}} \\
& & $\pi$  &   5.9 &   7.0 &   8.3 &   9.6 &  11.1 &  29.2 &  50.9 &  70.0 &  83.8 &\\
\cline{1-12}
{\multirow{2}{*}{$2$}} & {\multirow{2}{*} {$X^2$}} & $\alpha$ & \textit{5.6} &   \textit{5.5} &   5.1 &   5.2 &   5.1 &   5.1 &   5.1 &   5.1 &   5.0 &  \\
& & $\pi$ & \textit{6.0} &   \textit{7.0} &   8.4 &   9.8 &  11.1 &  30.0 &  52.2 &  71.2 &  85.0 &\\
 \hline
 \hline
 {\multirow{2}{*}{$3$}} & {\multirow{2}{*} {$X_{corr0}^2$}} & $\alpha$ &   5.4 &   5.3 &   5.2 &   5.1 &   5.1 &   5.0 &   5.0 &   5.1 &   5.0 & {\multirow{8}{*}{$\frac{p(x)}{g_2(x)}$}} \\
 &&$\pi$ &   6.0 &   6.7 &   7.8 &   8.8 &  10.0 &  25.0 &  43.9 &  61.5 &  76.2 &\\

 \cline{1-12}
 {\multirow{2}{*}{$4$}} & {\multirow{2}{*} {$X_{corr}^2$}} & $\alpha$                         &   3.9 &   4.4 &   4.6 &   4.7 &   4.7 &   5.0 &   5.0 &   5.0 &   4.9 &\\
 &&$\pi$ &   6.0 &   7.0 &   8.0 &   9.0 &  10.3 &  25.5 &  45.0 &  62.9 &  77.4 &\\

 \cline{1-12}
 {\multirow{2}{*}{$5$}} & {\multirow{2}{*} {$\hat X^2 $}} & $\alpha$                                   &   5.5 &   5.2 &   5.2 &   5.1 &   5.0 &   5.1 &   5.0 &   5.0 &   5.0 &\\
 &&$\pi$ &   6.1 &   7.1 &   8.1 &   9.2 &  10.6 &  26.3 &  46.0 &  64.1 &  78.5 &\\

 \cline{1-12}
 {\multirow{2}{*}{$6$}} & {\multirow{2}{*} {$ _{c}\hat X^2$}} & $\alpha$                                   &   5.1 &   5.0 &   5.0 &   5.0 &   5.0 &   5.0 &   5.0 &   5.0 &   5.0 &\\
 &&$\pi$ &   6.0 &   7.0 &   8.1 &   9.2 &  10.5 &  26.0 &  45.5 &  63.4 &  77.6 &\\

 \hline
 \hline
  {\multirow{2}{*}{$7$}} & {\multirow{2}{*} {$X_{corr0}^2$}} & $\alpha$                                             &  5.1 &   5.0 &  5.2  &   5.0 &   5.2 &   5.1 &   5.1 &   5.1 &   4.9 & \normalsize{\multirow{8}{*}{$\frac{p(x)}{g_3(x)}$}} \\
 &&$\pi$ &   6.3 &   7.5 &   8.8 &  10.7 &  12.3 &  35.3 &  60.5 &  79.6 &  91.3 &\\
 \cline{1-12}
 {\multirow{2}{*}{$8$}} & {\multirow{2}{*} {$X_{corr}^2 $}} & $\alpha$                                               & 3.5 &  4.1 & 4.5  &   4.7 &   4.8 &   5.0 &   5.0 &   5.0 &   4.9 &\\
 &&$\pi$ &   7.0 &   8.4 &   9.7 &  11.6 &  13.4 &  36.0 &  60.4 &  79.2 &  90.8 &\\
 \cline{1-12}
 {\multirow{2}{*}{$9$}} & {\multirow{2}{*} {$\hat X^2 $}} & $\alpha$                                                 & \textit{7.2} &  \textit{6.5} & \textit{6.0}   &   5.6 &   5.6 &   5.3 &   5.1 &   5.1 &   4.9 &\\
  &&$\pi$ &  \textit{16.4} &  \textit{30.1} &  \textit{40.1} &  48.8 &  56.1 &  95.7 &  99.8 & 100.0 & 100.0 &\\
 \cline{1-12}
 {\multirow{2}{*}{$10$}} & {\multirow{2}{*} {$_{c}\hat X^2$}} & $\alpha$                                                  &  \textit{4.6} &  \textit{ 4.9} &  \textit{4.9}  &   5.0 &   5.1 &   5.0 &   5.0 &   5.0 &   5.0 &\\
 &&$\pi$  &   \textit{7.0} &   \textit{8.5} &  \textit{10.0} &  11.7 &  13.5 &  37.0 &  61.7 &  80.0 &  91.2 &
\end{tabular}
\end{table}\\

Conclusion and  interpretation of results presented in Table \ref{tab:result25}.

\begin{itemize}
\item The size of all tests are close to nominal value 5\%.
\item The power of new tests $\hat X^2$(\ref{newx})(rows 5, 9) and $ _{c}\hat X^2$(\ref{unknown})(rows 6, 10) used for Poisson histograms  are greater than the power of tests  developed  ad hoc for the Poisson histograms $X_{corr0}^2$(\ref{stx15})(rows 3, 7) with the  exactly  known parameter $n_0$ and  $X_{corr}^2$(\ref{stx12})(rows 4, 8) with the unknown parameter $n_0$  in Ref. \cite{zech}.
\item The power of Pearson's test  $X^2$(\ref{basic11})(row 2) used for Poisson histograms is greater than test $X_{pois}^2$(\ref{basic})(row 1) with the exactly known parameter $n_0$  proposed in Ref. \cite{zech}.
\end{itemize}

\subsection{Numerical example 2}
A simulation study was done for the example described in Ref. \cite{zechebook} and also in Ref. \cite{gagpar}. Weighted histograms described in subsection 1.2 are used here.

  The PDF  $p_0(x)$ for the hypothesis $H_0$ is taken according to formula (\ref{p1_main})
with:
\begin{equation}
p_{0 tr}(x')=0.4(x'-0.5)+1; \,\, x'\in [0,1]
\end{equation}
\begin{equation}
A(x')=1-(x'-0.5)^2
\end{equation}
\begin{equation}
R(x|x')=\frac{1}{\sigma\sqrt{2\pi}}\exp\left[ -\frac{(x-x')^2}{2\sigma^2}\right], \text{with}\,\,\,   \sigma=0.3.
\end{equation}

For the alternative  $H_a$, $p(x)$ is taken with the same acceptance and resolution function according to formula (\ref{p1_main})
with:
\begin{equation}
p_{tr}(x')=0.6666(x'-0.5)+1; \,\, x'\in [0,1]
\end{equation}
that is presented by the weighted histogram.

\begin{figure}[t]
\vspace *{-0.0 cm}
\centering
\includegraphics[width=1 \textwidth]{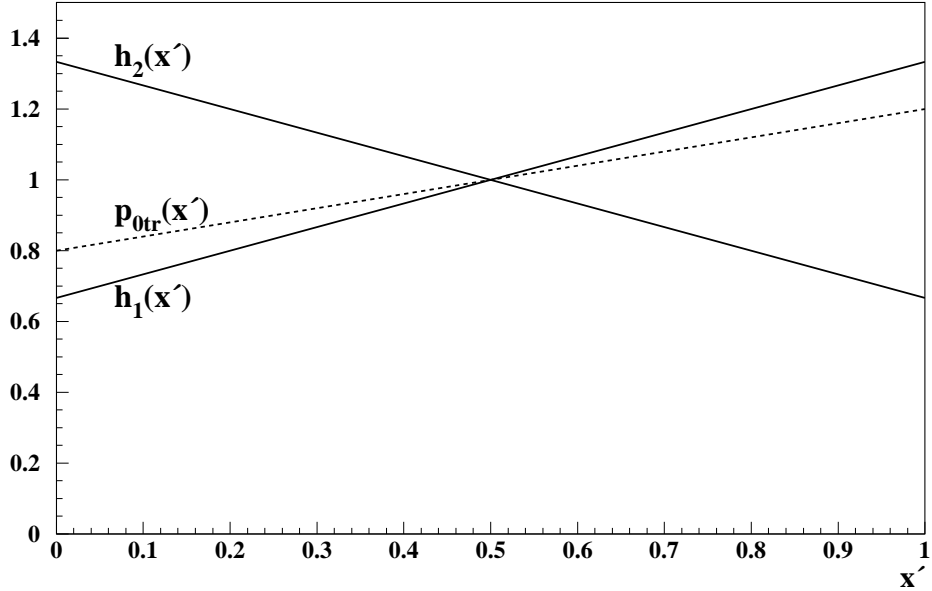}
\vspace *{-0.9cm}
 \caption {Probability density functions $h_1(x')=p_{tr}(x')$,  $h_2(x')$ and $p_{0tr}(x')$ (dashed line)}
 \label{fig:trhyal1}
\end{figure}

A calculation was done for two cases of PDFs used for event generation, see Fig.~\ref{fig:trhyal1}.
\begin{equation}
 h_1(x')=0.6666(x'-0.5)+1; \,\,\, x'\in [0,1]
 \end{equation}
 and\\
\begin{equation}
 h_2(x')=-0.6666(x'-0.5)+1; \,\,\, x'\in [0,1].
\end{equation}

\begin{figure}[h]
\centering
 \includegraphics[width=1 \textwidth]{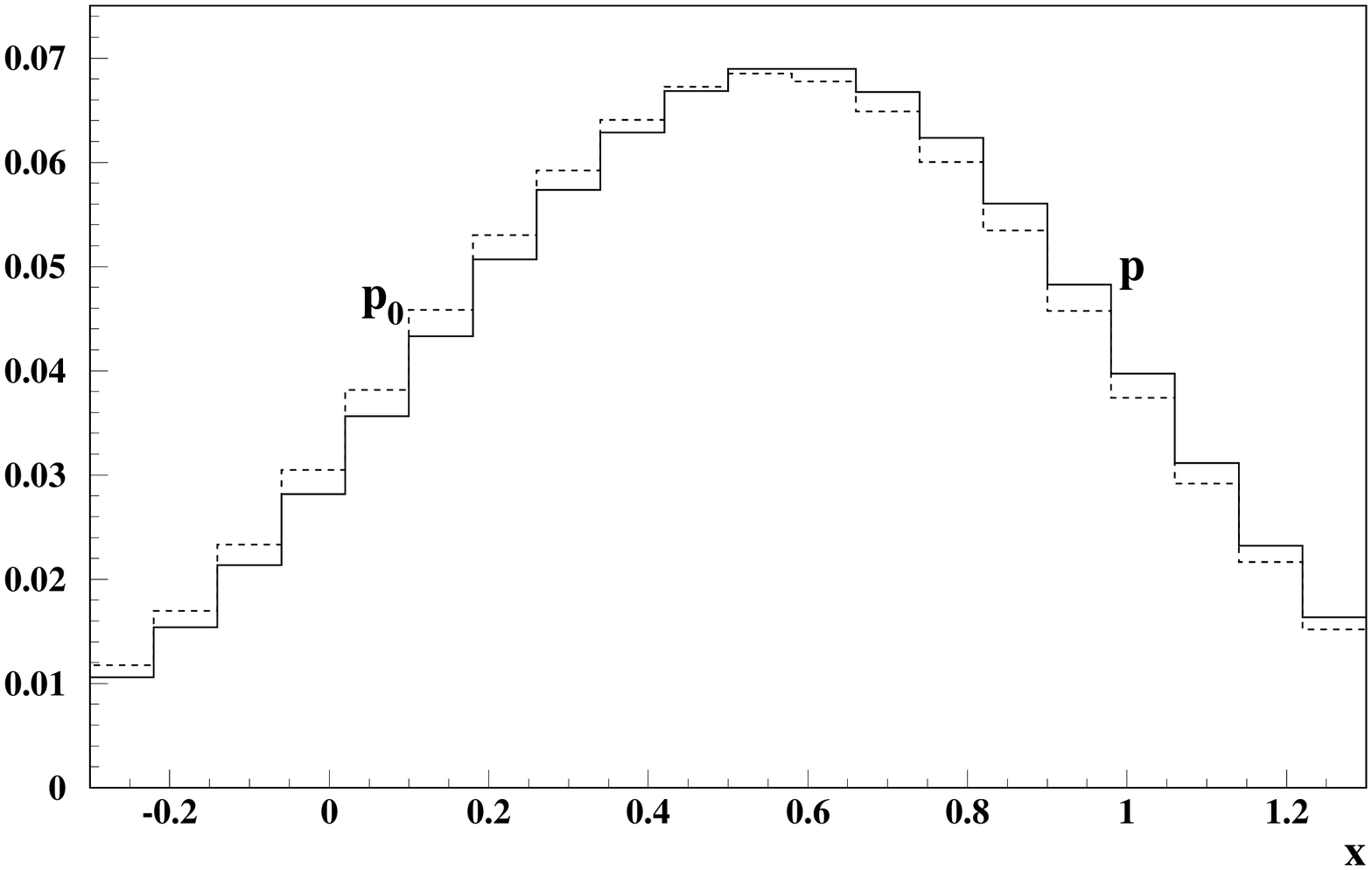}
  \caption {Probabilities $p_i, i=1,...,20$ for the PDF $p(x)$
(solid line)  and $p_{0i}, i=1,...,20$ for the PDF $p_{0}(x)$(dashed line)}
\label{fig:hyal}
\end{figure}

In the first case,  a weighted histogram is the histogram  with weights of events equal to 1 ( histogram with unweighted entries) and, in the second case, weights of events equal to $h_1(x')/h_2(x')$. The results of this calculation for 100000 runs are presented in tables 3. We use a histogram with 20 bins on interval $[-0.3,1.3]$.  Fig.~\ref{fig:hyal} presented probabilities  $p_i, i=1,...,20$ for the PDF $p(x)$ and $p_{0i}, i=1,...,20$ for the PDF $p_{0}(x)$.
 Here, we add two bins for events  with $x \leqslant -0.3$ and $x > 1.3$   as well as one bin for events that were not registered due to limited acceptance. Total number of bins $m$ is used in test  equal to 23.
The results of calculations of the sizes and power of tests for 100000 runs are presented in Table~\ref{tab:result82}.\\

 \begin{table}[h]
\centering
\caption{Numerical example 2. Sizes ($\alpha$) and powers ($\pi$) of different test statistics $X^2$(\ref{basic11}), $\hat X^2$(\ref{newx}), $_{c}\hat X^2$(\ref{unknown}), $\hat X_{Med}^2$(\ref{stdav}), $_{c}\hat X_{Med}^2$(\ref{stdav2}) obtained for different weighted functions $w(x)$. Italic type marks a size of test  with inappropriate number of events in the bins of  histograms.}
\label{tab:result82}
\vspace *{0.4 cm}
\scriptsize
\begin{tabular}{c|c|l|rrrrrrrrr|c}
\textnumero & & $n$ &   200 &   400 &   600 &   800 &  1000 &  3000 &  5000 &  7000 &  9000 & w(x) \\
 \hline
{\multirow{2}{*}{$1$}} & {\multirow{2}{*} {$X^2$}} & $\alpha$ & \textit{5.1} &   5.1 &   5.1 &   5.0 &   5.2 &   5.1 &   5.1 &   5.0 &   5.0 &{\multirow{2}{*}{$1$}} \\
& & $\pi$  &   \textit{5.6} &   6.6 &   7.5 &   8.8 &   9.8 &  25.9 &  45.7 &  64.9 &  79.4 &\\
 \hline
 \hline
 {\multirow{2}{*}{$2$}} & {\multirow{2}{*} {$\hat X^2$}} & $\alpha$ & \textit{  7.0} & \textit{6.2} &   5.8 &   5.6 &   5.5 &   5.1 &   5.0 &   4.9 &   4.9 & {\multirow{8}{*}{$\frac{h_1(x')}{h_2(x')}$}} \\
 &&$\pi$ &   \textit{8.4} &   \textit{9.4} &  10.9 &  12.8 &  14.6 &  40.9 &  67.1 &  85.3 &  94.5 &\\
 \cline{1-12}
 {\multirow{2}{*}{$3$}} & {\multirow{2}{*} {$_{c}\hat X^2$}} & $\alpha$                         &  \textit{ 5.6} & \textit{  5.6} &   5.5 &   5.4 &   5.3 &   5.1 &   5.0 &   5.0 &   4.9 &\\
 &&$\pi$ &   \textit{6.4} &   \textit{7.4} &   8.4 &   9.9 &  11.0 &  28.0 &  47.9 &  66.4 &  80.5 &\\

 \cline{1-12}
 {\multirow{2}{*}{$4$}} & {\multirow{2}{*} {$\hat X_{Med}^2 $}} & $\alpha$                                   & \textit{ 10.9} &  \textit{ 7.4} &   6.6 &   6.1 &   6.1 &   5.7 &   5.6 &   5.6 &   5.6 &\\
 &&$\pi$ &   \textit{9.1} &  \textit{10.1} &  11.5 &  13.9 &  15.8 &  43.7 &  70.9 &  87.8 &  95.8 &\\

 \cline{1-12}
 {\multirow{2}{*}{$5$}} & {\multirow{2}{*} {$ _{c}\hat X_{Med}^2$}} & $\alpha$                                   &  \textit{ 7.8} & \textit{  6.6} &   6.3 &   5.9 &   5.9 &   5.7 &   5.7 &   5.7 &   5.6 &
\\
 &&$\pi$ &   \textit{6.1} &   \textit{7.2} &   8.4 &   9.7 &  10.9 &  27.4 &  46.9 &  65.0 &  79.2 &
\end{tabular}
\end{table}

Conclusion and  interpretation of results presented in Table~\ref{tab:result82}.

\begin{itemize}
\item The size of new tests $\hat X^2$(\ref{newx}) and $ _{c}\hat X^2$(\ref{unknown}) (row 2, 3 ) is more close to the nominal value 5\%  then the size of median tests $\hat X_{Med}^2$(\ref{stdav}) and $_{c}\hat X_{Med}^2$(\ref{stdav2}) (rows 4, 5).
\item The power of new tests $\hat X^2$(\ref{newx})  and $_{c}\hat X^2$(\ref{unknown}) (rows 2,3)  is roughly the same compared with analogous median tests  $\hat X_{Med}^2$(\ref{stdav})  and  $_{c}\hat X_{Med}^2$(\ref{stdav2}) (rows 4, 5).
\item All tests demonstrate greater power then Pearson's test $X^2$(\ref{basic11}) (row 1) used for the histogram with unweighted entries.
\end{itemize}

The property of tests in application for Poisson histograms is investigated with the same weighted functions and distributions of events. In this case, the number of events $n$ in a histogram was simulated according Poisson distribution with given parameter $n_0$. The size and power of tests developed for the Poisson histogram in \cite{zech} was also  calculated.
Results of calculations are presented in Table~\ref{tab:result41}.
\begin{table}[h]
\centering
\caption{Numerical example 2. Size ($\alpha$) and power ($\pi$) of  different test statistics $X_{pois}^2$(\ref{basic}), $X^2$, $X_{corr0}^2$(\ref{stx15}), $X_{corr}^2$(\ref{stx12}), $\hat X^2$(\ref{newx}), $_{c}\hat X^2 $(\ref{unknown}) in application for Poisson histograms. Italic type marks a size of test  with inappropriate number of events in the bins of  histograms.}
\label{tab:result41}
\vspace *{0.4 cm}
\scriptsize
\begin{tabular}{c|c|l|rrrrrrrrr|c}
\textnumero & & $n_0$ &   200 &   400 &   600 &   800 &  1000 &  3000 &  5000 &  7000 &  9000 & w(x) \\
 \hline
 {\multirow{2}{*}{$1$}} & {\multirow{2}{*} {$X_{pois}^2$}}& $\alpha$ &   5.5 &   5.3 &   5.3 &   5.0 &   5.2 &   5.1 &   5.1 &   5.0 &   5.2 & {\multirow{4}{*}{$1$}}\\
& & $\pi$ &   5.5 &   6.4 &   7.4 &   8.7 &   9.7 &  25.6 &  45.0 &  64.0 &  77.9 &\\

\cline{1-12}
{\multirow{2}{*}{$2$}} & {\multirow{2}{*} {$X^2$}} & $\alpha$ & \textit{5.1} &   5.1 &   5.1 &   4.9 &   5.2 &   5.0 &   5.1 &   5.0 &   5.2 &
\\
& & $\pi$ &   \textit{5.5} &   6.5 &   7.5 &   8.9 &   9.8 &  26.3 &  45.9 &  65.0 &  78.8 &\\
 \hline
 \hline
 {\multirow{2}{*}{$3$}} & {\multirow{2}{*} {$X_{corr0}^2$}} & $\alpha$  &   5.8 &   5.8 &   5.5 &   5.3 &   5.3 &   5.0 &   5.1 &   5.0 &   5.0 & {\multirow{8}{*}{$\frac{h_1(x')}{h_2(x')}$}} \\
 &&$\pi$ &   5.8 &   6.6 &   7.7 &   9.0 &  10.2 &  27.2 &  47.0 &  66.5 &  80.7 &\\

 \cline{1-12}
 {\multirow{2}{*}{$4$}} & {\multirow{2}{*} {$X_{corr}^2$}} & $\alpha$                         &   4.2 &   4.9 &   4.9 &   4.8 &   4.9 &   4.9 &   5.0 &   4.9 &   5.0 &\\
 &&$\pi$ &   6.3 &   7.2 &   8.4 &   9.7 &  11.1 &  27.5 &  46.9 &  65.5 &  79.5 &\\

 \cline{1-12}
 {\multirow{2}{*}{$5$}} & {\multirow{2}{*} {$\hat X^2 $}} & $\alpha$                                   &  \textit{ 6.8} &  \textit{ 6.1} &   5.8 &   5.6 &   5.4 &   5.1 &   5.0 &   5.0 &   4.9 &\\
 &&$\pi$ &   \textit{8.4} &   \textit{9.4} &  11.0 &  12.7 &  14.8 &  41.1 &  67.2 &  85.1 &  94.4 &\\

 \cline{1-12}
 {\multirow{2}{*}{$6$}} & {\multirow{2}{*} {$ _{c}\hat X^2$}} & $\alpha$ &   \textit{5.4} &  \textit{ 5.6} &   5.5 &   5.3 &   5.3 &   5.0 &   4.9 &   5.0 &   4.9 &                                  \\
 &&$\pi$   &   \textit{6.5} &   \textit{7.4} &   8.5 &   9.8 &  11.0 &  28.0 &  48.2 &  66.4 &  80.4 &
\end{tabular}
\end{table}

Conclusion and  interpretation of results presented by Table~\ref{tab:result41}.

\begin{itemize}
\item The size of all tests are close to nominal value 5\%.
\item Basically, the power of new tests $\hat X^2$(\ref{newx})  and $ _{c}\hat X^2$(\ref{unknown})(rows 5, 6) in applying   for Poisson histograms  are greater than the power of tests  developed ad hoc for the Poisson histograms $X_{corr0}^2$(\ref{stx15}) with the exactly known parameter $n_0$ and  $X_{corr}^2$(\ref{stx12}) (rows 3, 4) with  the unknown parameter $n_0$ in Ref. \cite{zech}.
\item The power of Pearson's test  $X^2$ (row 2) used for Poisson histograms is greater than power of test $X_{pois}^2$(\ref{basic})(row 1) with the exactly known parameter $n_0$.
\end{itemize}

Generally the numerical example 1 and example 2 demonstrate  the superiority of new goodness of fit tests under existing tests for weighted histograms, see Ref. \cite {gagunash} and for weighted Poisson histograms, see Ref. \cite{zech}.
\newpage

\section{Conclusion}

A review of goodness of fit tests for weighted histograms was presented.
The bin content of a weighted histogram  was considered as a random sum of random variables that  permits to generalize  the classical Pearson's goodness of fit test for  histograms with weighted entries.
 Improvements of  the chi-square tests with better statistical properties were proposed. Evaluation of the size and power of  tests was done numerically for different types of weighted histograms with  different numbers of events and different weight functions. Generally the size of new tests is  closer to  nominal value  and power is not lower than have  existing tests. Except direct application of tests in data analysis, see for example Ref. \cite {physlet}, the proposed tests are necessary  bases for generalization of  test in the case when  some parameters  must be estimated from the data, see Ref. \cite{cramer},  as well as for the generalisation of test for  comparing  weighted and unweighted histograms  or two weighted ones (homogeneity test), see Refs. \cite {cramer,gagcomp,gagcpc}. Parametric fit  of data obtained from detectors with finite resolution and limited acceptance is one of important  application of  methods developed for weighted histograms  that can be used for  experimental data interpretation, see Refs. \cite {gagpar}.
\\

{\bf{Acknowledgements}}
\\
\noindent

The author is grateful to Johan Blouw for useful
 discussions and careful reading of the manuscript and thanks the University of Akureyri and the MPI for Nuclear Physics for support in carrying out the research.

\end{document}